%% file: sample-sigconf.tex
\documentclass[sigconf]{acmart}

\usepackage{booktabs} 
\usepackage{todonotes}
\usepackage{algorithm}
\usepackage{algpseudocode}
\usepackage{setspace}
\usepackage{subcaption}
\usepackage{dirtytalk}
\usepackage{pifont}
\usepackage{nicefrac}

\setcopyright{none}

\acmDOI{10.475/123_4}

\acmISBN{123-4567-24-567/08/06}

\acmConference[AdKDDTargetAd '18]{2018 AdKDD \& TargetAd Workshop
in conjunction with
The 24th ACM SIGKDD Conference on Knowledge
Discovery and Data Mining}{Aug 2018}{London, UK} 
\acmYear{2018}
\copyrightyear{2018}

\acmPrice{15.00}

\begin{document}
\title{Designing Experiments to Measure Incrementality on Facebook}

\author{C. H. Bryan Liu}
\orcid{0000-0002-6516-2364}
\affiliation{%
  \institution{ASOS.com}
  \city{London} 
  \country{United Kingdom}
}
\email{bryan.liu@asos.com}

\author{Elaine M. Bettaney}
\affiliation{%
  \institution{ASOS.com}
  \city{London} 
  \country{United Kingdom}
}
\email{}

\author{Benjamin Paul Chamberlain}
\affiliation{%
  \institution{Imperial College London \& ASOS.com}
  \city{London} 
  \country{United Kingdom}
}
\email{}



\renewcommand{\shortauthors}{C. H. B. Liu et al.}

\begin{abstract}

The importance of Facebook advertising has risen dramatically in recent years, with the platform accounting for almost 20\% of the global online ad spend in 2017. 
An important consideration in advertising is incrementality: how much of the change in an experimental metric is an advertising campaign responsible for. 
To measure incrementality, Facebook provide lift studies. As Facebook lift studies differ from standard A/B tests, the online experimentation literature does not describe how to calculate parameters such as power and minimum sample size. 
Facebook also offer multi-cell lift tests, which can be used to compare campaigns that don't have statistically identical audiences. In this case, there is no literature describing how to measure the significance of the difference in incrementality between cells, or how to estimate the power or minimum sample size.
We fill these gaps in the literature by providing the statistical power and required sample size calculation for Facebook lift studies. 
We then generalise the statistical significance, power, and required sample size calculation to multi-cell lift studies.
We represent our results theoretically in terms of the distributions of test metrics and in practical terms relating to the metrics used by practitioners, making all of our code publicly available. 
\end{abstract}

%
%



\maketitle

\input{samplebody-conf}

\bibliographystyle{ACM-Reference-Format}
\bibliography{testing-bibliography} 

\end{document}

%% file: samplebody-conf.tex
\section{Introduction}

In 2017, advertisers spent \$204bn online~\cite{zenith18advertising}, with a large share (\$40bn) spent targeting Facebook's 2.13bn monthly active users~\cite{facebook18facebook}.
To maximise their return on investment, advertisers continuously test and optimise their campaigns. It is increasingly common to use controlled experiments to maximise the incrementality of an advertising campaign. In the most common variant --- known as A/B, or split testing --- the target population is divided into two groups, a test group, where members are shown adverts, and a control group, where members are not shown adverts. The difference in a metric of interest (e.g. total sales or number of app installs) between the test group and control group is the \emph{incrementality} of the campaign. Facebook offers advertisers the opportunity to measure the incrementality of their campaigns via \emph{lift studies}.

Despite the importance of Facebook advertising, there is a lack of literature or documentation describing how to design experiments. The deficiencies are summarised in Table~\ref{tbl:existing_literature}.\footnote{On their experimentation website~\cite{facebook18whatmakes}, Facebook state that ``To build a study with more rigorous calculations, or for more information on Conversion or Brand Lift, please reach out to your Facebook Account Representative.''} We address this issue by first describing how Facebook calculate incrementality and using this to derive measures of  statistical significance, the test power and the minimum sample size for Facebook lift studies.

\begin{table}
  \begin{tabular}{ l | c | c }
  \centering
      Existing literature on & Lift studies & Multi-cell lift studies \\
      \hline
      Test statistic & \cite{gordon2017comparison} & \ding{55} \\
      Statistical significance & \cite{gordon2017comparison} & \ding{55} \\
      Power / Required sample size & \ding{55} & \ding{55} \\
  \end{tabular}
  \vspace*{5pt}
\caption{Existing literature on calculating the test statistic (lift/incrementality), its statistical significance, test power, and the required sample size for Facebook lift studies and multi-cell lift studies. The only literature available is the white paper by Gordon et al.~\cite{gordon2017comparison}.}
  \vspace*{-15pt}
\label{tbl:existing_literature}
\end{table}

A Facebook lift study is similar to an A/B test with two important differences. Firstly, the control group is scaled so that the size of the test and control groups are the same. This changes the variance of the metric of interest in the control group.\footnote{If the control group is scaled up, the variance increases. Likewise the variance decreases if the control group is scaled down.} 
Secondly, not everyone in the test group is shown an advert. This happens because the advertiser can lose every bid for a particular user, or when a bid is won, the advert appears off the screen. Members of the test group who are shown the advert at least once during the test period are referred to as the \emph{reached audience}, and those who have not seen the advert during the test period are referred to as the \emph{unreached audience}. The activity of the unreached audience introduces variance that is not present in a standard A/B test, which must be factored in when calculating the power and required sample size. 

Facebook has a mechanism that takes the scaled control group and the unreached audience into account when reporting on the incrementality and its associated statistical significance~\cite{gordon2017comparison} (see Section~\ref{sec:lift_study}), but they do not cover the statistical power or required sample size. We introduce these calculations in this paper.

\begin{figure}
\begin{center}
    \includegraphics[width=0.48\textwidth, trim=0 0 20mm 0, clip]{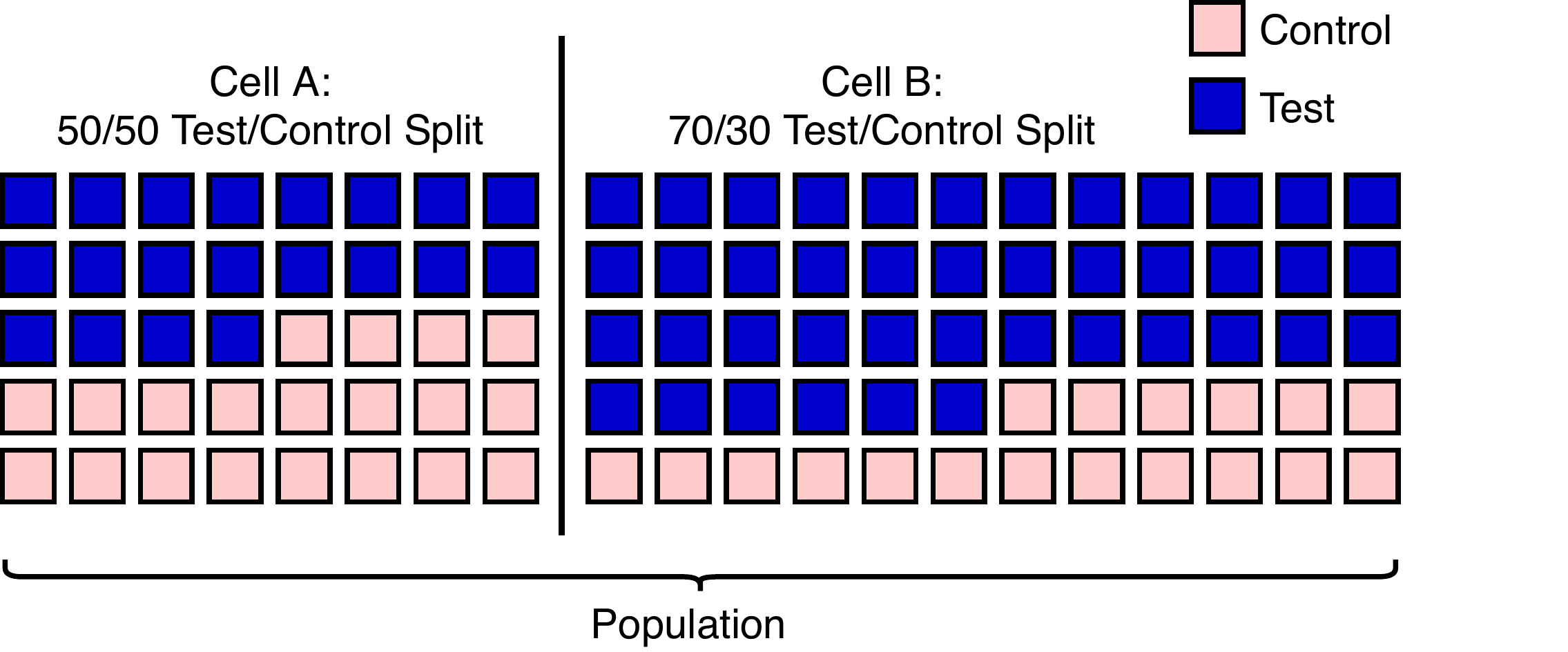}
\end{center}
\caption{A Facebook multi-cell lift study. The population (100 boxes), is randomly divided into multiple cells. Different campaigns with differing test-control splits can be run in each cell.}
\label{fig:multi_cell_illustration}
\vspace{-10mm}
\end{figure}

Facebook also support \emph{multi-cell lift studies}, where the target population is split into multiple \emph{cells} each with a control and test group of their own, as illustrated in Figure~\ref{fig:multi_cell_illustration}. These can be used to compare two marketing strategies where the target audience exhibits a selection bias~\cite{liu2018online}. An example is comparing campaigns that vary the bid size based on customer lifecycle, which result in a different user composition between the cells. In this case we are interested in measuring the difference between incrementalities attained by the campaigns.

While Facebook reports the incrementality of each individual cell in a multi-cell lift study, they do not report if the incrementality difference is statistically significant, nor advise on the statistical power or sample size required to design the experiment. A common pitfall is to apply the standard sample size calculation for a lift study to a multi-cell lift study. As there are more test/control groups in a multi-cell experiment, the variance of the test metric will be larger, even when the groups have the same size. Furthermore, changes in marketing strategies are likely to lead to changes in audience composition meaning that test group metrics from multiple cells are not directly comparable via standard t-tests.  Permutation tests are also not possible in this setting as Facebook do not provide data regarding the control-test split.

We resolve these problems by introducing a framework to calculate the power and minimum sample size for lift studies and multi-cell lift studies on Facebook. Our framework takes into account control group scaling and the effect of the unreached audience. 
We present our calculations both theoretically and in practical terms. Our theoretical results relate to the distribution of the test metrics, while in practical terms, we present results in the metrics used by advertising practitioners (e.g. lift or proportion of reached audience).

To summarise, our contributions are:
\begin{enumerate}
    \item We derive the statistical power and required sample size for Facebook lift studies, bridging the gap between the online controlled experimental literature and the reality on measuring incrementality on Facebook.
    \item We generalise the results to multi-cell lift studies, where incrementalities under different strategies are compared against each other.
    \item We make our result useful to advertising practitioners by presenting our statistical power and minimum required sample size calculations in terms of expected lift, reach percentage, and the ratio between test/control groups, as well as making the code used in the paper publicly available.\footnote{\url{https://github.com/liuchbryan/fb_lift_study_design}}
\end{enumerate}

In the remainder of the paper we derive the distribution of the test metric and hence the test power and minimum sample size required in a Facebook lift study in Section~\ref{sec:lift_study}. We then generalise the results to multi-cell lift studies in Section~\ref{sec:multicell_lift_study}. Finally, we show a number of empirical results illustrating the correctness of the derived distributions and the difference in the required sample sizes in single-cell/multi-cell lift studies in Section~\ref{sec:evaluation}.

\section{Facebook Lift Studies}
\label{sec:lift_study}

We first describe a lift study, concentrating on how Facebook derives the incrementality and lift (relative incrementality) of the metric of interest in Section~\ref{sec:fb_incrementality_calculation}. We then base our derivation of the distribution of lift as a test statistic (Section~\ref{sec:lift_distribution_derivation}), as well as calculations on the test power and required samples size (Section~\ref{sec:lift_study_power}) on their work. We will use conversions, defined as the number of transactions from users in the lift study, as our metric of interest, but our calculations are applicable to other metrics which can be described with a Poisson process.\footnote{For metrics which cannot be described with a Poisson process, our framework, which supports the use of a simulated distribution generated from arithmetic operations of samples drawn from Poisson distributions, can still be applied by swapping in different base distributions.}

\subsection{How does Facebook calculate incrementality and lift?}
\label{sec:fb_incrementality_calculation}

Facebook manages the test-control splitting and is therefore able to measure the conversions in each group.  Facebook reports three results: (1) the number of conversions in the test group $C_T$, (2) the number of conversions in the control group $C_C$ and (3) the number of conversions from the reached audience in the test group $R_T$.  The sizes of the test and control groups are also reported enabling the control group to be scaled to match the total audience of the test group. We base our calculations on the conversions in the control group, which is scaled so that the audience size matches that in the test group:
\begin{align}
	C_S = s C_C \;,
    \label{eq:scaling_factor_def}
\end{align}
where $s$ is the ratio of the test to control group sizes
\begin{align}
	s = \frac{N_T}{N_C}.
\end{align}

\begin{figure}
\begin{center}
    \includegraphics[width=0.35\textwidth]{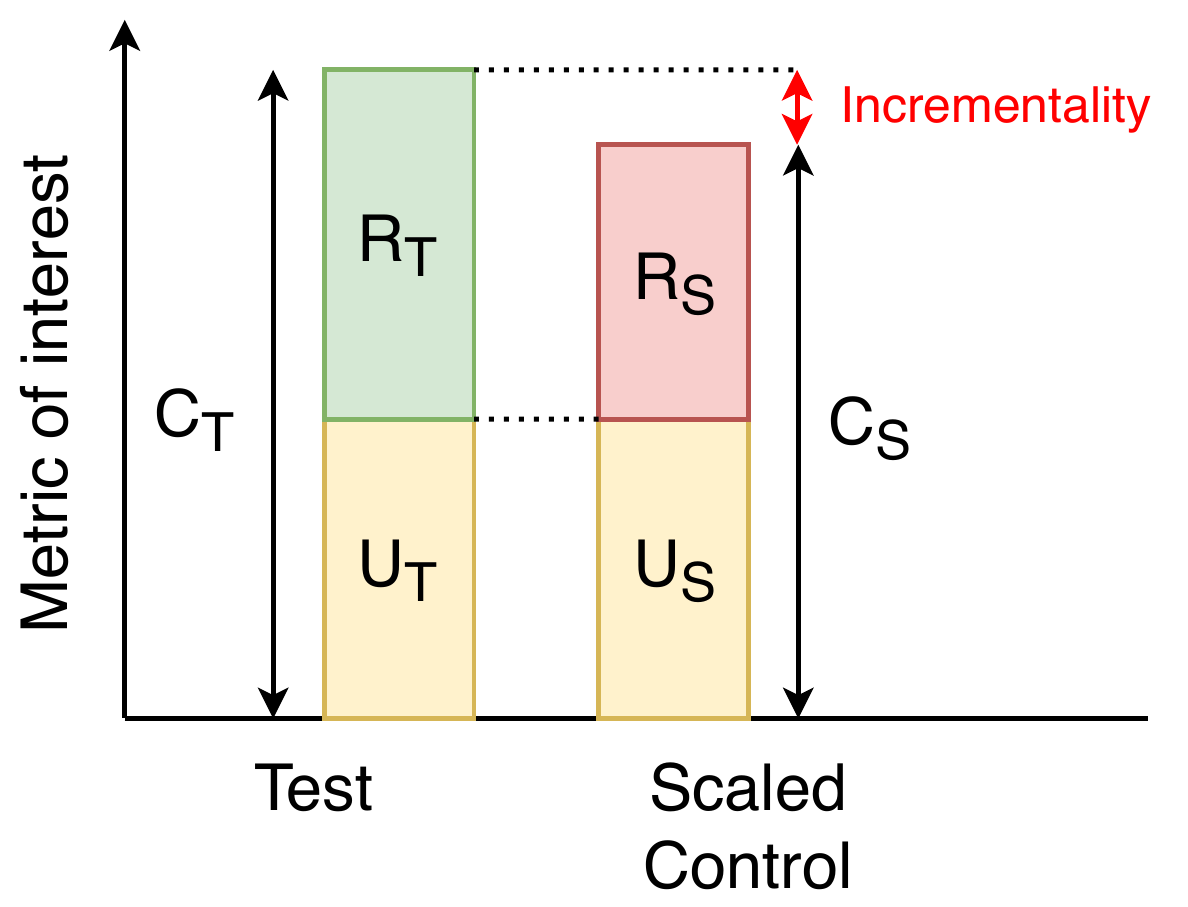}
\end{center}
\caption{The Facebook incrementality calculation. $C_T$ and $C_S$ represent the metric attained by the test and scaled control groups respectively. $R_T$ and $R_S$ represent the contribution by the reached audience in the test and scaled control groups respectively. $U_T$ and $U_S$ represent the contribution of the unreached audience in the test and scaled control groups respectively.}
\label{fig:FB_lift_study}
\end{figure}

The conversions in the test and scaled control groups contain contributions from both the reached $R$ and unreached $U$ audiences
\begin{align}
	C_T = R_T + U_T\,, \quad C_S = R_S + U_S\;,
\end{align}
and these are illustrated in Figure~\ref{fig:FB_lift_study}.  Since the conversion rates in both unreached audiences are assumed to be the same
\begin{align}
	U_S = U_T.
\end{align}

Reach $r$ is defined as the fraction of people in the test group who saw an advert
\begin{align}
	r = \frac{N_{T_R}}{N_T},
\end{align}
where $N_{T_R}$ is the size of the reached audience and $N_T$ is the total audience size of the test group.  We assume that the reach would be the same in both test and control groups, hence
\begin{align}
	r = \frac{N_{C_R}}{N_C},
\end{align}
where $N_{C_R}$ is the size of the audience who \textit{would} have been shown an advert in the control group. In the control group the conversion rates are the same in the unreached and reached audiences and so
\begin{align}
	r = \frac{R_C}{C_C} = \frac{R_S}{C_S}.
    \label{eq:reach_def}
\end{align}

The incrementality is the difference in conversions between the test and scaled control groups and originates solely from the reached audiences
\begin{align}
	I = C_T - C_S = R_T - R_S \;.
\end{align}

\label{sec:lift_study_test_statistic}
The test statistic is lift ($L$) defined as incrementality divided by the number of reached conversions in the scaled control
\begin{align}
	L = \frac{C_T - C_S}{R_S},
    \label{eq:lift_def}
\end{align}
which can be calculated in terms of $C_T$, $C_C$ and $R_T$ as
\begin{align}
	L = \frac{C_T - s\,C_C}{s\,C_C - C_T + R_T}.
    \label{eq:lift_def1}
\end{align}

Facebook's Null Hypothesis Significance Test determines if there is a non-zero lift at 90\% confidence level (two-tailed). In our calculations, we focus on the alternate hypothesis that a campaign is incremental at 5\% significance level (one-tailed).\footnote{While the calculations around test power and required sample size is nearly identical in both formulations, we are assuming an advert will not have a negative incrementality. This is most often the case when we run control experiments to measure an advert's incrementality.} Formally
\begin{align}
	\textrm{H}_0 : \mathbb{E}(L) = 0, \quad \textrm{H}_1 : \mathbb{E}(L) > 0 \;,
\end{align}
where $\textrm{H}_0$ is the null and $\textrm{H}_1$ the alternate hypothesis.

\subsection{Derivation of the lift distributions}
\label{sec:lift_distribution_derivation}
To obtain the power and required sample size for a lift study, it is necessary to understand the distributions of the test statistic under the null and alternate hypotheses. Here we derive the distribution of the test statistic $L$, which is not available in the literature.\footnote{We take $L$ as the relative difference between a Poisson variable and the \emph{scalar multiple} of a Poisson variable. This rules out the use of the Poisson means test~\cite{KRISHNAMOORTHY200423}, which compares two standard Poisson variables with potentially different rates.} We begin by observing that $R_S$ is defined to be a scalar multiple of $C_S$ by Equation~\eqref{eq:reach_def}, and hence $L$ can be written as
\begin{align}
    L = \frac{C_T}{R_S} - \frac{C_S}{R_S} = \frac{C_T}{R_S} - \frac{1}{r},
    \label{eq:lift_minus_constant_form}
\end{align}
where $r$ is the reach. We assume $C_T$ follows a Poisson distribution with rate $\lambda_T$, and $R_S$ is $C_C$, an independent Poisson random variable with rate $\lambda_C$, scaled by a factor of $rs$ (i.e. $R_S = rs\cdot C_C$, by Equations~\eqref{eq:reach_def} and~\eqref{eq:scaling_factor_def}). The probability mass functions (PMF) of $C_T$ and $R_S$ is then given as:
\begin{align}
   & f_{C_T}(x) = e^{-\lambda_T} \frac{\lambda_T^x}{x!}\; ,\; x \in \mathbb{N} ;\\
   & f_{R_S}(x) = f_{C_C}\left(\frac{x}{rs}\right) = e^{-\lambda_C} \frac{\lambda_C^{(x/rs)}}{(x/rs)!}\; ,\; x \in \left\{0, {rs}, 2{rs}, ...\right\} = {rs}\mathbb{N} ,
    \label{eq:R_S_distribution}
\end{align}
where Equation \eqref{eq:R_S_distribution} is a standard result on transformation of univariate random variables.

The cumulative mass function (CMF) of $L$ is
\begin{align}
    F_L(l) & = \mathbb{P}(L \leq l) = \mathbb{P}\left(\frac{C_T}{R_S} - \frac{1}{r} \leq l\right)
           = \mathbb{P}\left(\frac{C_T}{R_S} \leq l + \frac{1}{r} \right) \\
           & \approx \mathbb{P}\left(C_T \leq \left(l + \frac{1}{r}\right) R_S \right)\;,
\end{align}
where we use approximately equal in the expression as the probability distribution of $C_T/R_S$ is not well defined.\footnote{$R_S$ can be equal to zero, leading to the quotient having an undefined value with positive probability. In practice, with $\lambda_C$ being sufficiently large (say over 30, achieved by a sufficient number of naturally occurring conversions) we can safely proceed as the probability of~$R_S$ equal to zero is negligible (${\mathbb{P}(R_S = 0 \,|\, \lambda_C = 30) < 10^{-13}}$ and the probability decreases with increasing $\lambda_C$). Alternatively, we can model $C_C = \nicefrac{1}{rs} (R_S)$ as a zero-truncated Poisson distribution, though with all these random variables related to each other by some arithmetic operations, this approach will introduce other complications when deriving the distribution of~$L$.} The CMF has the form
\begin{align}
    F_L(l) & \approx \sum_{i \in {rs}\mathbb{N}} \sum_{j=0}^{\lfloor (l + 1/r) i \rfloor} f_{C_T}(j)\,f_{R_S}(i)\;.
\end{align}
The outer summation is difficult to implement as it is defined over $rs\mathbb{N}$, and $rs$ is unknown a priori. We substitute $k = \nicefrac{i}{(rs)}$ so that the outer summation sums over the natural numbers and uses the PMF of $C_C$ instead (see Equation~\eqref{eq:R_S_distribution}):
\begin{align}
    F_L(l) & \approx \sum_{k=0}^{\infty} \sum_{j=0}^{\lfloor (l + 1/r)(rs)\cdot k \rfloor} f_{C_T}(j)\,f_{C_C}(k) \\
    & = \sum_{k=0}^{\infty} \sum_{j=0}^{\lfloor (l + 1/r)(rs)\cdot k \rfloor} e^{-(\lambda_T + \lambda_C)} \frac{\lambda_T ^j \, \lambda_C^k}{j! \, k!} \;,\; l \in \mathbb{Q}.
    \label{eq:L_distribution}
\end{align}

The derived distribution can then be used to calculate the critical value of $L$, above which $\textrm{H}_0$ should be rejected. The critical value is necessary for calculating the power and required sample size.

\subsection{Power and Minimum Sample Size Calculation}
\label{sec:lift_study_power}
A prerequisite of any A/B test is a calculation of the expected test power and the minimum sample size to achieve an acceptable test power.\footnote{Typically taken to be 0.8 .} While we have derived the necessary CMF to calculate power and sample size, we also explore the possibility to proceed by simulating the distribution for $L$ using a large number of samples. We show in Section~\ref{sec:derived_simulated_comparison} that the derived and simulated distributions are equivalent, and there are computational advantages to using the simulation approach. The simulation is also applicable if we assume the variables used in this section follow other distributions.

\subsubsection{Power}
Test power is the probability that the test will correctly reject the null hypothesis $\textrm{H}_0$ when the alternate hypothesis $\textrm{H}_1$ is true (the complement of Type II error). 
For Facebook lift studies, test power is dependent on 
the minimum detectable lift~$L_m$, the number of expected conversions in the control group~$\mathbb{E}(C_C)$, the scaling factor relating the size of the test group to the control group~$s$, and the reach~$r$, which depends on many variables, in particular ad spend.

To calculate the test power we require the distribution for~$L$. This can be done by  using Equation~\eqref{eq:L_distribution}. Alternatively, we can obtain an empirical distribution for $L$ by 1) treating~$C_C$ and $C_T$ as Poisson random variables with means $\lambda_{C}$ and $\lambda_{T}$ respectively, 2) drawing samples from $C_C$ and $C_T$, and using Equations~\eqref{eq:reach_def} and~\eqref{eq:scaling_factor_def} to scale them to obtain samples for $R_S$ and $C_S$, and 3) using Equation~\eqref{eq:lift_def} to obtain samples for $L$.

We calculate the means $\lambda_{C}$ and $\lambda_{T}$ by expressing them in terms of $\mathbb{E}(C_C)$, $r$ and expected lift $\mathbb{E}(L)$.
We can approximate $\lambda_{C}$ with
\begin{align}
	\lambda_{C} = \mathbb{E}(C_C)\;, 
\end{align}
and are then able to calculate $\lambda_{T}$ as
\begin{align}
	\lambda_{T} = \mathbb{E}(C_T) = s \lambda_{C}  (1 + r \,\mathbb{E}(L))\;,
    \label{eq:lam_c_t}
\end{align}
by rearranging Equation~\eqref{eq:lift_minus_constant_form} and noting the scaling relationship between $R_S$ and $C_C$ using Equations~\eqref{eq:reach_def} and~\eqref{eq:scaling_factor_def}.



The procedure for calculating the test power is two-fold and is illustrated in Figure~\ref{fig:sc_dist}.  First, the distribution of $L$ is calculated under $\mathrm{H_0}$ in which $\lambda_{T} = s\,\lambda_{C}$ (i.e. $\mathbb{E}(L) = 0$).  Estimates for  $\mathbb{E}(C_C)$ and $r$ can be taken from previous Facebook advertising results.  For a one-tailed test at the 5\% significance level the critical value $c$ is calculated as the 95th percentile of this distribution:
\begin{align}
    F_L(c \,|\, \textrm{H}_0 \textrm{ is true}) = \mathbb{P}(L \leq c \,|\, \mathbb{E}(L) = 0) = 0.95 \;.
\end{align}
Second, the distribution of $L$ is calculated under a specific ${\mathrm{H_1}: \mathbb{E}(L) = L_m}$ in which $\lambda_{T}$ is as defined in Equation~\eqref{eq:lam_c_t}. Since the test power is strongly coupled to $L_m$ (see Figure~\ref{fig:power_sim}), it is important to have a reasonable estimate. Estimates for $L_m$ can be taken from previous Facebook advertising results. If no previous studies are available, we can estimate $L_m$ from a lightweight pre-study, or related studies in the literature.  The test power $1 - \beta$ can then be calculated as the percentage of this distribution above $c$:
\begin{align}
   1 - \beta & = \mathbb{P}(L > c \,|\, \mathbb{E}(L) = L_m) \\
   & = 1 - \mathbb{P}(L \leq c \,|\, \mathbb{E}(L) = L_m) 
   = 1 - F_L(c \,|\,  \textrm{H}_1 \textrm{ is true}) \;.
\end{align}

\subsubsection{Minimum sample size}
\label{sec:min_sample_size}
The minimum sample size required to give a specified test power $p$ (commonly 80\%) can be obtained from the power simulation by solving for the minimum $\mathbb{E}(C_C)$ that will give a power greater than $p$ using the bisection method~\cite{burden1985numerical}. The minimum sample sizes to observe lifts of 1\%, 2\%, 5\% and 10\% are shown in Table~\ref{tbl:sample_size}.

\begin{table}
  \begin{tabular}{ l | r | r | r | r }
  \centering
      & \multicolumn{2}{l |}{Single-cell} & \multicolumn{2}{l}{Multi-cell} \\
      Effect size & $C_C$ & $N$ & $C_{C,A}$ & $N$ \\
      \hline
      10\% & 1,352 & 54,068 & 2,745 & 219,596 \\
      5\% & 5,107 & 204,271 & 10,754 & 860,346 \\
      2\% & 31,571 & 1,262,848 & 67,453 & 5,396,260 \\
      1\% & 124,459 & 4,978,355 & 264,745 & 21,179,569 \\
  \end{tabular}
  \vspace*{5pt}
\caption{Minimum number of conversions in the control group $C_C$ and total audience size $N$ required to achieve a power of 80\%. For the multi-cell calculation the lift in cell A was taken to be 5\%. To calculate the total audience size, we divide $C_C$ by the conversion rate\protect\footnotemark~(assumed to be 5\%), and multiply the result by the number of groups (two for single-cell, and four for two-cell lift studies).}
\label{tbl:sample_size}
  \vspace*{-15pt}
\end{table}

\footnotetext{Defined as the number of conversions divided by the total number of users.}

\section{Multi-cell lift studies}
\label{sec:multicell_lift_study}
Multi-cell lift studies can be used to compare the incrementalities of multiple marketing strategies with potentially statistically different audiences.  Here we consider the case of two cells, $A$ and~$B$. To maximise the test power, we assume the cells are of the same size, with the same test-control split proportions.
A common pitfall in multi-cell studies is to use the test power and minimum sample size derived in Section~\ref{sec:lift_study}. As multi-cell studies have more test/control groups, the variance of the test statistic, which involves arithmetic operations on all groups, will increase even if the variance within each group stays the same. In Section~\ref{sec:comparison_single_multi_cell} we demonstrate this and develop the mechanism for correctly calculating test parameters. 

In a multi-cell lift study, Equations~\eqref{eq:lift_def} and \eqref{eq:lift_def1} still hold for individual cells:
\begin{align}
    L_A = \frac{C_{T,A} - C_{S,A}}{R_{S,A}}\,, \quad
    L_B = \frac{C_{T,B} - C_{S,B}}{R_{S,B}} \;,
    \label{eq:multicell_lift_def_original}
\end{align}
where the additional subscripts $A$ and $B$ indicate the cells.
Facebook provide advertisers with $C_{T,A}$, $C_{C,A}$, $R_{T,A}$, $C_{T,B}$, $C_{C,B}$ and $R_{T,B}$ so $L_A$ and $L_B$ can be computed as
\begin{align}
L_A = \frac{C_{T,A} - s\,C_{C,A}}{s\,C_{C,A} - C_{T, A} + R_{T,A}}\,, \quad
L_B = \frac{C_{T,B} - s\,C_{C,B}}{s\,C_{C,B} - C_{T, B} + R_{T,B}} \;.
    \label{eq:multicell_lift_def}
\end{align}

\paragraph{Test Statistic}
We define the test statistic as the \emph{absolute} (as opposed to relative) difference between the lifts in cells $A$ and $B$:
\begin{align}
	D = L_B - L_A \;,
    \label{eq:multicell_lift_difference_def}
\end{align}
which is directly comparable with the lift in a single-cell study.\footnote{If we define the test statistic as the relative difference, the effect size between cells will be a percentage of the effect size achieved in the single-cell case. To illustrate, a 1\% relative difference in lifts means we are comparing a 5\% lift in cell A and a 5.05\% lift in cell B. To detect such difference with 80\% power we require around 106M conversions in the control group of cell A (one out of four groups in a two-cell lift study), a number which even the largest companies struggle to meet for experimentation purposes.} The null and alternative hypotheses are defined to be
\begin{align}
	\textrm{H}_0 : \mathbb{E}(D) = 0, \quad \textrm{H}_1 : \mathbb{E}(D) > 0.
\end{align}

While the distributions for $L_A$ and $L_B$ can be characterised by their CMF, it is difficult to obtain the PMF of these distributions. Accordingly, the distribution of $D$ (e.g. the CMF $F_D(\cdot)$ or PMF $f_D(\cdot)$) can not be readily evaluated using a convolution. 
We believe that deriving an analytical form for the distribution of $D$ is of little practical use for test power and sample size calculation as there are other simpler alternatives such as simulating the distribution.

Under $\textrm{H}_0$ the distribution of $D$ is defined by $r$, $s$, $\mathbb{E}(L_A)$, $\mathbb{E}(C_{C,A})$ and~$\mathbb{E}(C_{C,B})$.  It is reasonable to assume that $r$ and $s$ are the same for both cells.  In general, the audiences are not statistically identical in cells $A$ and $B$ so that $C_{C,B}=C_{C,A}$ can not be assumed.  However, if the strategy in $B$ has not previously been tested, there is no good way of estimating $C_{C,B}$ and so we assume $C_{C,B}=C_{C,A}$ here.

\paragraph{Statistical Significance \& Critical Value}
As Facebook do not report the difference in lifts between cells (or its significance) in multi-cell studies, advertisers are free to choose the significance level $\alpha$ that suits their needs. We use a one-tailed test at 5\% for the calculations shown in Section~\ref{sec:comparison_single_multi_cell} to be consistent with Section~\ref{sec:lift_study}.

The critical value $c$
is defined to satisfy the following equation:
\begin{align}
    F_D(c \,|\, \textrm{H}_0 \textrm{ is true}) = 1 - \alpha \;.
\end{align}
This can be obtained by finding the $100(1-\alpha)$ percentile of the samples simulating the distribution of $D$.


\paragraph{Power}
  Under $\textrm{H}_1$ we define a minimum detectable difference $D_m$ such that
\begin{align}
	\mathbb{E}(L_B) = \mathbb{E}(L_A) + D_m \,,
\end{align}
and calculate the test power $1-\beta$ by the following equation:
\begin{align}
   1 - \beta 
   = 1 - F_D(c \,|\,  \mathbb{E}(D) = D_m) \;,
\end{align}

\paragraph{Minimum sample size}
The minimum sample sizes required to be able to observe $D_m = 1\%, 2\%, 5\%, 10\%$ with a power of 80\% were calculated as described in Section \ref{sec:min_sample_size}. The equivalent numbers of conversions in cell $A$ control and total audience sizes are shown in Table~\ref{tbl:sample_size}. 

\section{Evaluation}
\label{sec:evaluation}

In this section, empirical results on the distribution of the test statistic in single-cell lift studies and the calculated power and sample size in both single-cell and multi-cell lift studies are provided. In Section~\ref{sec:derived_simulated_comparison} we show the correctness of our simulation of $L$ by comparing it to the analytical form in Equation~\eqref{eq:L_distribution}. Finally, in Section~\ref{sec:comparison_single_multi_cell}, we calculate the test power and required sample size for a range of minimum detectable effects, for both single-cell and multi-cell lift studies. 

\subsection{Comparing the derived and simulated distribution of $L$}
\label{sec:derived_simulated_comparison}

\begin{figure}
\captionsetup[subfigure]{justification=centering}
\centering
	\begin{subfigure}{0.23\textwidth}
    	\centering
		\includegraphics[width=\textwidth, trim=1mm 0 1mm 0, clip]{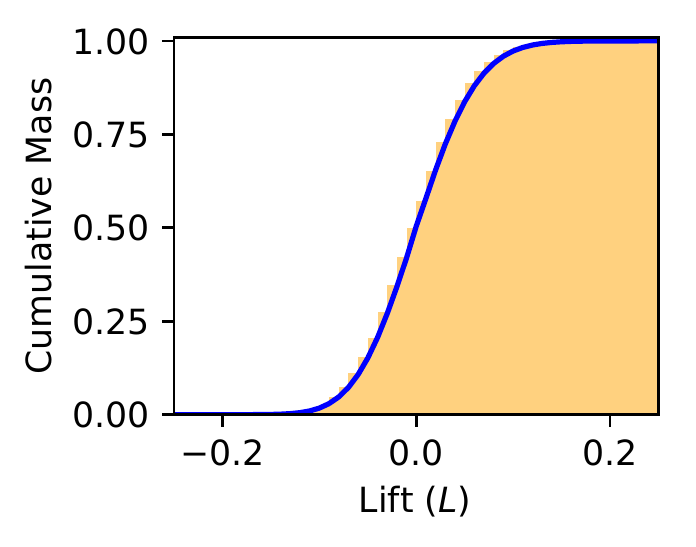}
        \vspace*{-15pt}
      	\caption{$\lambda_T = 1000$, $\lambda_C = 1000$,\\ $r = 1$, $s=0.9$}
        \label{fig:cmf_empirical_comparison_1}
    \end{subfigure}
    \begin{subfigure}{0.23\textwidth}
    	\centering
		\includegraphics[width=\textwidth, trim=1mm 0 1mm 0, clip]{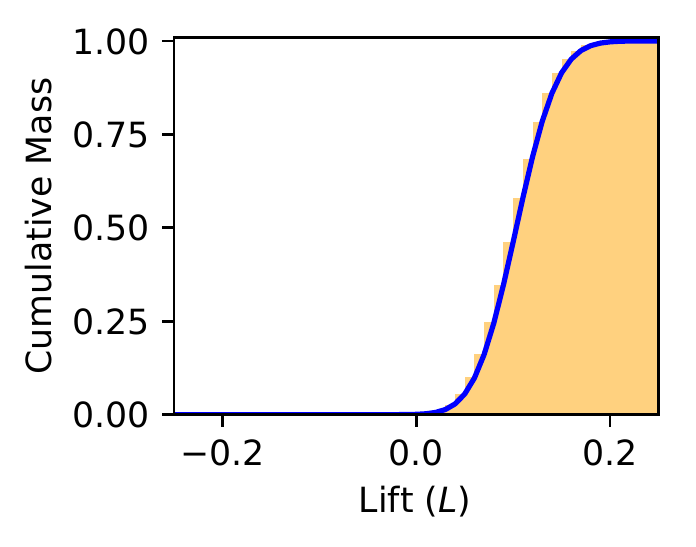}
        \vspace*{-15pt}
      	\caption{$\lambda_T = 4644$, $\lambda_C = 7189$,\\ $r = 0.6088$, $s=0.5916$}
        \label{fig:cmf_empirical_comparison_2}
    \end{subfigure}
    \begin{subfigure}{0.23\textwidth}
    	\centering
		\includegraphics[width=\textwidth, trim=1mm 0 1mm 0, clip]{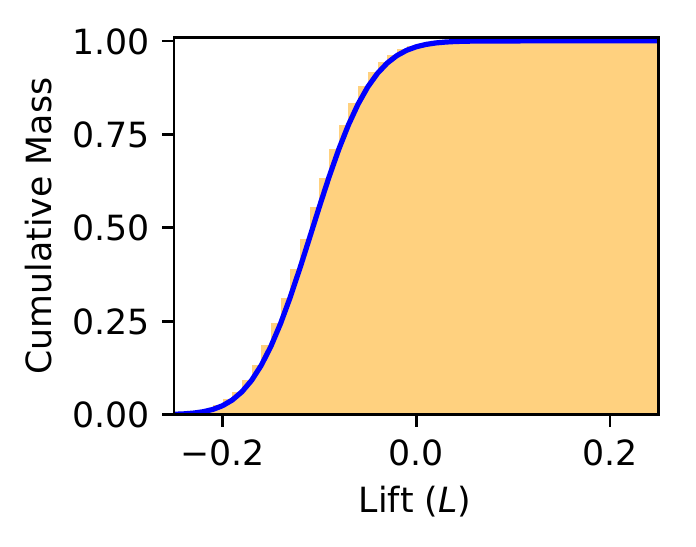}
        \vspace*{-15pt}
      	\caption{$\lambda_T = 3745$, $\lambda_C = 3009$,\\ $r = 1.3121$, $s=0.4812$}
        \label{fig:cmf_empirical_comparison_3}
    \end{subfigure}
    \begin{subfigure}{0.23\textwidth}
    	\centering
		\includegraphics[width=\textwidth, trim=1mm 0 1mm 0, clip]{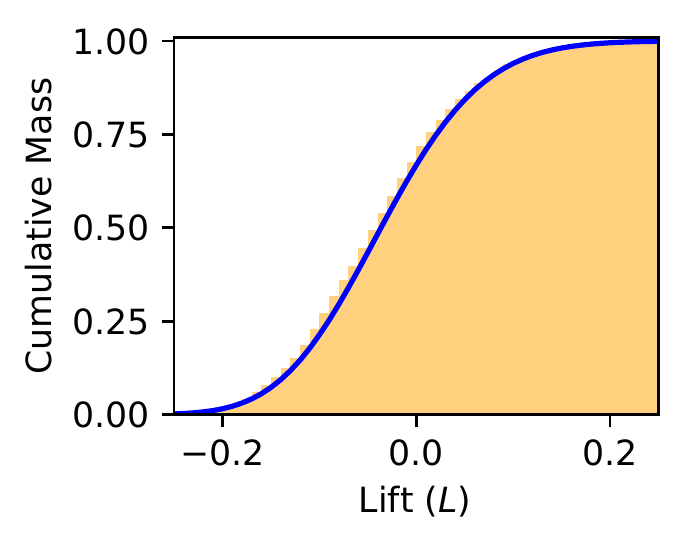}
        \vspace*{-15pt}
      	\caption{$\lambda_T = 2297$, $\lambda_C = 408$,\\ $r = 5.7654$, $s=0.6361$}
        \label{fig:cmf_empirical_comparison_4}
    \end{subfigure}
    \vspace*{-5pt}
    \caption{Comparison between the CMF of the lift derived in Section~\ref{sec:lift_distribution_derivation} (blue line) and the cumulative histogram of  1,000 samples drawn from the generative process in Section~\ref{sec:lift_study_power} (orange bars). Over a large range of the parameters $\lambda_T$, $\lambda_C$, $r$, and $s$, the two methods produce largely identical distributions.
    }
    \vspace*{-5pt}
    \label{fig:cmf_empirical_comparison}
\end{figure}

We first confirm that our simulation of $L$ (specified in Equation~\eqref{eq:L_distribution}) is correct by running a number of Kolmogorov-Smirnov (K-S) tests~\cite{smirnov1944approximate,daniel1978applied}. This indicates that the simulated distribution can be safely used as an alternative for the purpose of power and required sample size calculation.

For each run we 1)~randomly specify the four parameters required by both methods: $\lambda_T$, $\lambda_C$, the reach $r$, and the scaling factor $s$, 2)~generate a number of samples from the simulated distribution, 3)~compute the K-S statistic w.r.t. the derived distribution, and 4)~evaluate if there are any statistical significance to reject the null hypothesis that the two distributions are the same. Steps 3) and 4) are mostly handled by the \texttt{kstest} function in \texttt{scipy}.

We had 500 test runs (four are shown in Figure~\ref{fig:cmf_empirical_comparison}), and 28 of them have a K-S statistic that results in rejecting the null hypothesis at a 5\% significance level. Taking into account that we are running multiple comparisons and hence should expect around 25 rejections given the two distributions are the same, we are satisfied that the derived and simulated distributions are statistically equivalent.

It is more than 30 times quicker to obtain the 95th percentile of the distribution of $L$ (i.e. the critical value) using the simulated distribution than the derived distribution. This is done by comparing the time taken to:
\begin{itemize}
    \item (Simulated distribution) Find the value of the 95th percentile in the 10M samples simulating the distribution, versus
    \item (Derived distribution) Find the root of the function ${F_L(l)- 0.95}$ under the same parameters, using the root-finding algorithm proposed by Brent~\cite{brent2013algorithms}.
\end{itemize}
This suggests it is more effective for an advertiser to obtain the test power using the simulated distribution for the single-cell case.

\subsection{Comparison of single-cell and multi-cell test power and minimum sample size}
\label{sec:comparison_single_multi_cell}

\begin{figure}
\centering
	\begin{subfigure}{0.236\textwidth}
    	\centering
		\includegraphics[width=\textwidth, trim=2.5mm 0 2.5mm 0, clip]{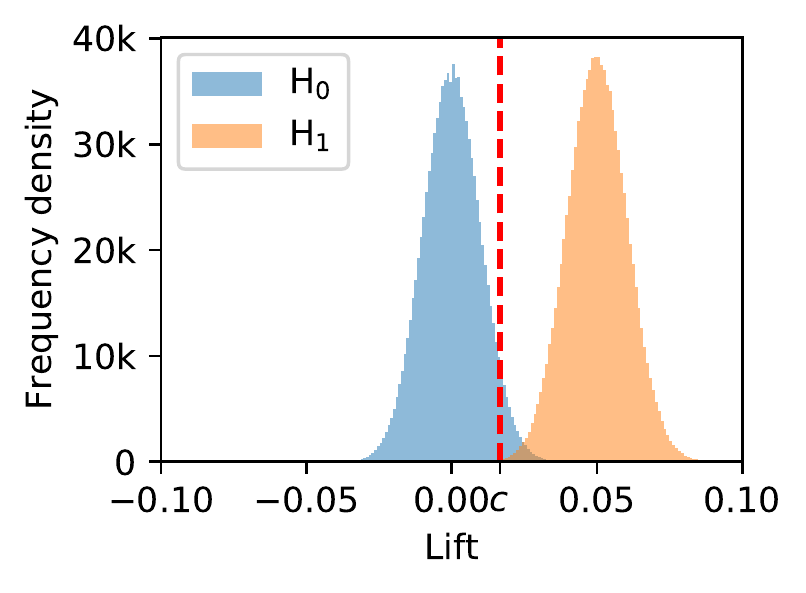}
        \vspace*{-15pt}
      	\caption{}
        \label{fig:sc_dist}
    \end{subfigure}
    \begin{subfigure}{0.236\textwidth}
    	\centering
		\includegraphics[width=\textwidth, trim=2.5mm 0 2.5mm 0, clip]{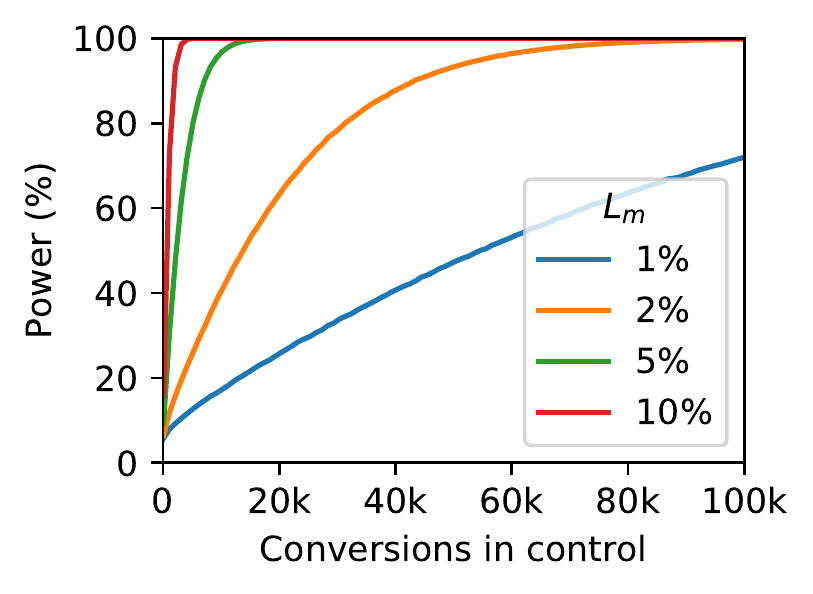}
        \vspace*{-15pt}
      	\caption{}
        \label{fig:sc_power_vs_conv}
    \end{subfigure}
    \begin{subfigure}{0.236\textwidth}
    	\centering
		\includegraphics[width=\textwidth, trim=2.5mm 0 2.5mm 0, clip]{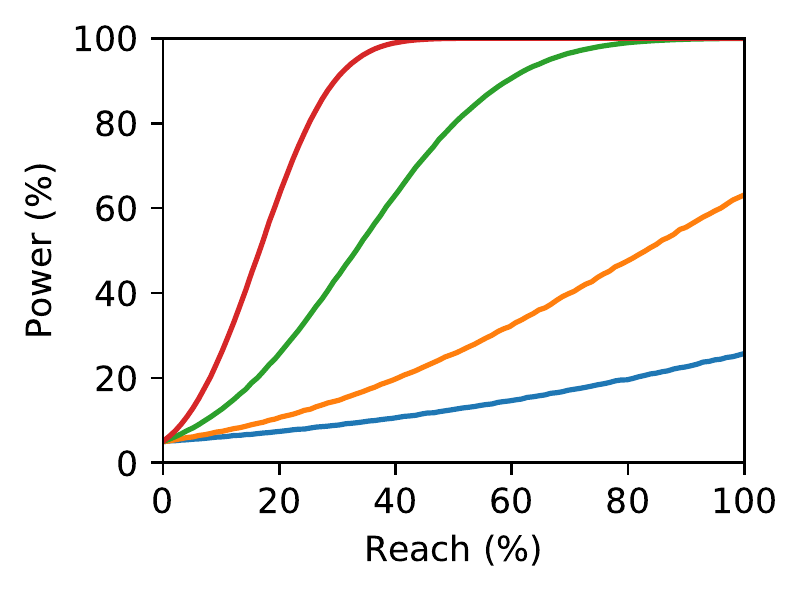}
        \vspace*{-15pt}
      	\caption{}
        \label{fig:sc_power_vs_reach}
    \end{subfigure}
    \begin{subfigure}{0.236\textwidth}
    	\centering
		\includegraphics[width=\textwidth, trim=2.5mm 0 2.5mm 0, clip]{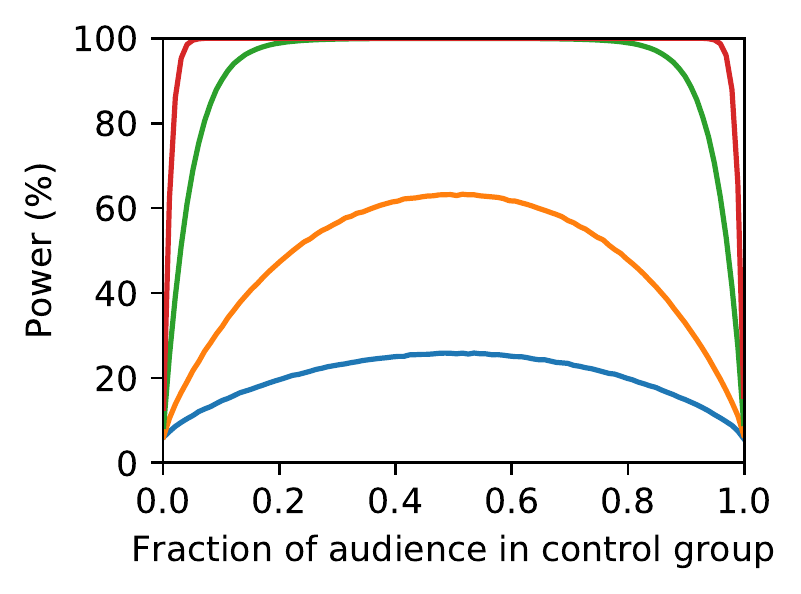}
        \vspace*{-15pt}
      	\caption{}
        \label{fig:sc_power_vs_split}
    \end{subfigure}
    \begin{subfigure}{0.236\textwidth}
    	\centering
		\includegraphics[width=\textwidth, trim=2.5mm 0 2.5mm 0, clip]{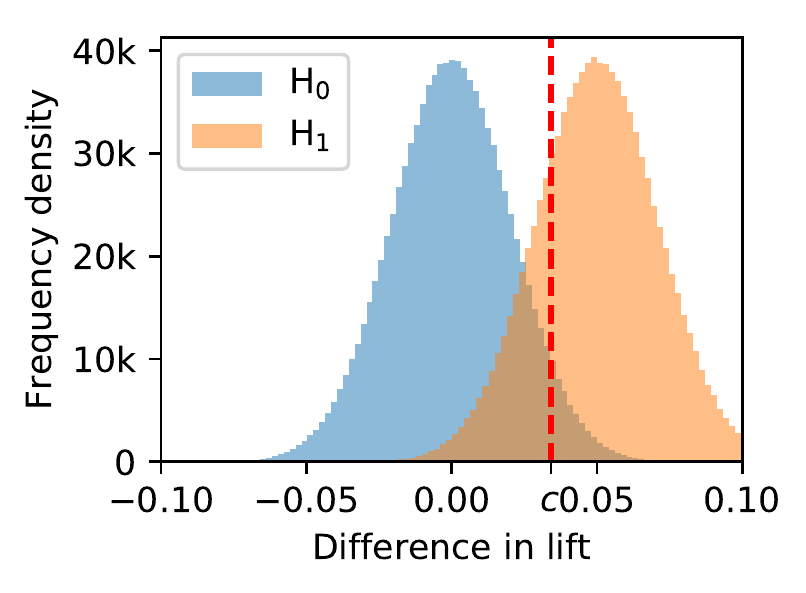}
        \vspace*{-15pt}
      	\caption{}
        \label{fig:mc_dist}
    \end{subfigure}
    \begin{subfigure}{0.236\textwidth}
    	\centering
		\includegraphics[width=\textwidth, trim=2.5mm 0 2.5mm 0, clip]{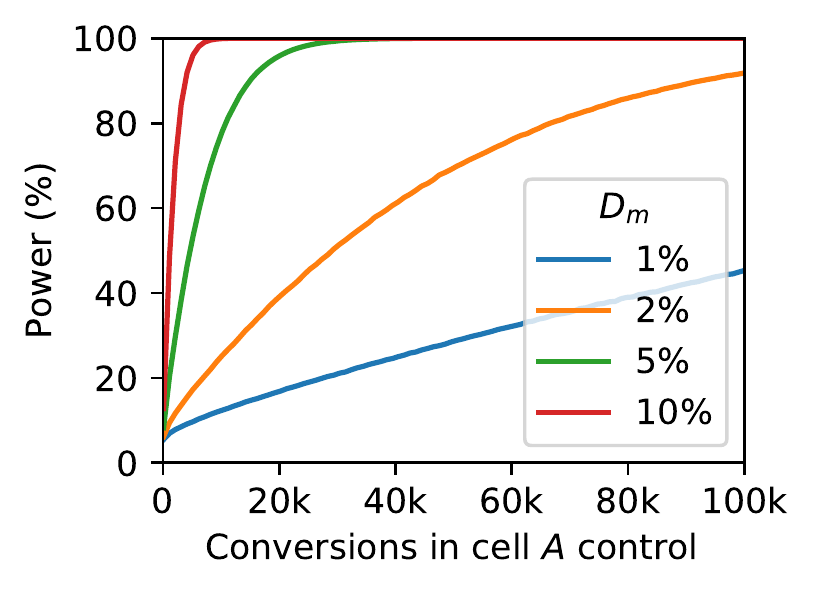}
        \vspace*{-15pt}
      	\caption{}
        \label{fig:mc_power_vs_conv}
    \end{subfigure}
    \vspace*{-5pt}
    \caption{Simulations for single-cell (a-d) and multi-cell (e-f) lift studies. a)~Distributions of $L$ under $\mathrm{H_0}$ and $\mathrm{H_1}$ for 20,000 conversions in the control group, true lift of 5\%, reach of 100\% and a 50:50 control-test split. $c$ marks the critical value for a one-tailed test at the 5\% significance level. b)~Test power against the number of control conversions for different minimum detectable lifts. c)~Test power against reach percentage holding the total audience size constant ($C_C = 20\mathrm{k}$). d)~Test power against the fraction of audience in the control group, holding the total audience size constant ($C_C = 20\mathrm{k}$ when the test/control split is 50:50) e)~Distributions of the difference in lift between two cells under $\mathrm{H_0}$ and $\mathrm{H_1}$ where the true difference is 5\%. f)~Test power against the number of conversions in the control group for different minimum detectable relative differences in lift.}
    \label{fig:power_sim}
    \vspace*{-10pt}
\end{figure}

Finally, we visualise our power and required sample size calculations, recording the number of conversions (and thus users) required to detect certain effects in both single-cell and multi-cell lift studies. 

Figures~\ref{fig:power_sim}a \& e show the power calculation for the single and multi-cell cases respectively.  To be comparable, the total audience size $N$ is fixed s.t. $C_C = 20\mathrm{k}$ and $C_{C,A} = 10\mathrm{k}$.
The power in the multi-cell case of 78\% (with $D_m=5\%$) is meaningfully lower than the 100\% power achieved in the single-cell case (with $L_m=5\%$).
Figures~\ref{fig:power_sim}b, c \& d show the variation of single-cell test power with audience size, reach and control-test split respectively.  For a given audience size the maximum power can be obtained with a reach of 100\% and a 50:50 split between the test and control groups (where~$s=1$). 
Figure~\ref{fig:mc_power_vs_conv} is the multi-cell equivalent of Figure~\ref{fig:sc_power_vs_conv}.  Comparing these figures shows that for the same number of conversions per control group, the power achieved is less in the multi-cell case.  Furthermore, this effect is larger for smaller effect sizes.

Table~\ref{tbl:sample_size} shows that to achieve a test power of 80\% over twice as many conversions are needed per control group in the multi-cell than in the single-cell case.  Since our multi-cell scenario has two cells, the total audience size needed in the multi-cell is over four times that of the single-cell case.

\section{Conclusion}
\label{sec:conclusion}
We have described how to design experiments to measure the incrementality of advertising campaigns on Facebook, bridging the gap between the general literature in online controlled experiments and industrial practices.
We provided the statistical power and required sample size calculation for Facebook lift studies, and generalised the statistical significance, power and required sample size calculation to multi-cell lift studies, which are used by advertisers to compare campaigns or strategies where the target audience can exhibit a selection bias.
We make our results useful to practitioners by presenting our calculations in terms of common advertising metrics --- expected lift, reach percentage, and ratio between test/control groups --- and publishing all of our code.

\begin{acks}
The authors thank Markus Ojala and Lauri Kovanen for useful discussions and the anonymous reviewers for providing many improvements to the original manuscript.
\end{acks}